\documentclass[prd,eqsecnum,notitlepage,
nofootinbib,superscriptaddress]
{revtex4-1}

\global\arraycolsep=2pt
\usepackage{stmaryrd}
\usepackage{amsmath}
\usepackage{amssymb}
\usepackage{graphicx}
\usepackage{textcomp}
\usepackage{calrsfs}
\usepackage{yfonts}
\usepackage{bm}
\usepackage{color}
\newcommand{\ben}{\begin{equation*}}
\newcommand{\een}{\end{equation*}}
\newcommand{\bean}{\begin{eqnarray*}}
\newcommand{\eean}{\end{eqnarray*}}

\newcommand{\nn}{\nonumber}
\newcommand{\be}{\begin{equation}} 
\newcommand{\ee}{\end{equation}}
\newcommand{\bea}{\begin{eqnarray}} 
\newcommand{\eea}{\end{eqnarray}}

\DeclareMathOperator{\sgn}{sgn}

\begin{document}
\title{Electrodynamic friction of a 
charged particle passing a conducting plate}
\author{Kimball A. Milton}
  \email{kmilton@ou.edu}
  \affiliation{H. L. Dodge Department of Physics and Astronomy,
University of Oklahoma, Norman, OK 73019, USA}
\author{Yang Li}
\email{liyang@ou.edu}
 \affiliation{H. L. Dodge Department of Physics and Astronomy, University of 
Oklahoma, Norman, OK 73019, USA} 
\author{Xin  Guo} 
\email{guoxinmike@ou.edu}  
\affiliation{H. L. Dodge Department of Physics and Astronomy, University of 
Oklahoma, Norman, OK 73019, USA}
\author{Gerard Kennedy}\email{g.kennedy@soton.ac.uk}
\affiliation{School of Mathematical Sciences,
University of Southampton, Southampton, SO17 1BJ, UK}

\begin{abstract}
The classical electromagnetic friction of a charged particle, moving 
with prescribed constant velocity $\mathbf{v}$ parallel to a planar
imperfectly conducting surface, is reinvestigated.  As a concrete example, the 
Drude model
is used to describe the conductor. The transverse electric and transverse
magnetic contributions have very different characters both in the low-velocity 
(nonrelativistic)
and high-velocity (ultrarelativistic) regimes.  Both numerical and analytical 
results are given.  Most remarkably, the transverse magnetic contribution
to the friction has a maximum for $|\mathbf{v}|<c$, and
 persists in the limit of vanishing resistivity for
sufficiently high velocities.
 We also show how Vavilov-\v{C}erenkov radiation can
be treated in the same formalism.
\end{abstract}

\date\today
\maketitle

\section{Introduction}
\label{sec:intro}
Over the past several decades, there has been continuing theoretical interest 
in Casimir
or quantum friction between dielectric bodies in relative motion, or between 
polarizable
atoms and dielectric or conducting surfaces, but, to date, there has been 
no experimental confirmation of
such effects.  For a brief review with many references, see 
Ref.~\cite{Milton:2015aba}.

In the course of our continuing investigations, we have also examined 
classically analogous
effects.  For example, a charged particle moving close to an imperfectly 
conducting surface
experiences a drag force parallel to its motion.  This was apparently first 
considered by Boyer
\cite{boyer1974}, and later revisited  
\cite{boyer1996,tw1997,tomassone1997}.  Ohmic heating is the relevant
physical mechanism \cite{sokoloff}, and the phenomenon may have been
observed in experiments with solid nitrogen sliding above (superconducting)
lead \cite{dayo, renner, krim}, although, in such a case, quantum effects
are likely to be more relevant \cite{persson}.

Here, we will extend these nonrelativistic studies into the relativistic 
regime,
 continuing to model the conductor by a Drude-type dispersion relation, 
and analyze the very different behaviors of the
transverse electric (TE) and transverse magnetic (TM) contributions.
The physical origin of the friction in the classical and quantum regimes is
the same---the dissipation in the surface---so understanding this better 
in the classical case may yield useful insight into the quantum case.

Of course, it will be recognized that, for real metals, the Drude model is 
only appropriate for $\hbar\omega \alt 1$ eV \cite{olmon}. 
Therefore, our work should be 
regarded mainly as an illustrative theoretical exercise. However, since 
the same methods can be generalized in a straightforward manner to a more 
appropriate description of an imperfect conductor, we would expect our results 
and conclusions to remain qualitatively correct in that context.


The outline of this paper is as follows.  In Sec.~\ref{sec:gen}, we set up
our general formulation in terms of TE and TM Green's functions.  The TE
contribution is discussed in Sec.~\ref{sec:TE}, with analytical results
for both low and high velocities, while a similar treatment for the
somewhat more complex, but more important,
 TM contribution is given in Sec.~\ref{sec:TM}. 
In the latter case, we find very interesting
nonmonotonic effects, as well as the persistence
of friction in the limit of vanishing damping.
A brief discussion of possibilities of observing such effects is given
in Sec.~\ref{sec:concl}.  Appendix \ref{appa} provides more
detail on the electromagnetic Green's functions, while Appendix \ref{appc}
shows how analytical expressions for integrals encountered in
intermediate- and  high-velocity regimes
 are obtained.  Appendix \ref{sec:vc} demonstrates 
how the same formulation
can be used to describe the motion of a charged 
particle in a uniform dielectric
medium, and the force on the particle due to Vavilov-\v{C}erenkov 
radiation is rederived.

In this paper, we will use Heaviside-Lorentz units with $c=1$.

\section{General expressions}
\label{sec:gen}
The idea is very simple.  A particle of charge $e$ is moving 
with velocity $\mathbf{v}$ parallel to a 
plane conducting surface.  It experiences the Lorentz force
\be
\mathbf{F}=e(\mathbf{E+v\times B}).
\ee
The magnetic field
does no work on the particle, so may be disregarded.  
The electric field arises because of the image charge induced by the 
conducting plane.
This field may be expressed in terms of a suitable Green's dyadic, 
 most conveniently written in the frequency domain:
\be
\mathbf{E(r};\omega)=-\frac1{i\omega}\int (d\mathbf{r}') 
\bm{\Gamma}(\mathbf{r,r'};\omega)\cdot\mathbf{j(r'};\omega).\label{egf}
\ee
(For the connection with the perhaps more familiar Green's function expressed 
in terms of
vector potentials, see the Appendix of Ref.~\cite{Schwinger:1977pa}. 
For further details, see Appendix \ref{appa}.) 
Here, the current is due to the
particle moving with prescribed constant velocity 
$\mathbf{v}=v\mathbf{\hat x}$, parallel to, and at 
 a distance $a$ in the $z$ direction above, the surface of the conductor:
\be
\mathbf{j(r},t)=e\mathbf{v}\delta(\mathbf{r}-a\mathbf{\hat z}
-\mathbf{v}t)=ev\mathbf{\hat x}\delta(x-vt)\delta(y)\delta(z-a).
\ee

We choose this representation for the Green's dyadic because it is precisely 
the retarded version of that used
in the quantum calculations that are the main focus
 of our research on friction.  Because the conductor 
lies in the $x$-$y$
plane, we have translational invariance in that plane, which permits the 
transverse Fourier transform,
\be
\mathbf{\Gamma(r,r'};\omega)=\int\frac{(d\mathbf{k}_\perp)}{(2\pi)^2}
e^{i\mathbf{k_\perp\cdot
(r-r')_\perp}}\mathbf{g}(z,z';\mathbf{k}_\perp,\omega), \quad k^2
=\mathbf{k}_\perp^2.\label{tft}
\ee
Inserting this construction into the Lorentz force formula, we immediately 
obtain the frictional force along the direction of motion,
\be
F=F_x=-\frac{e^2}{2\pi i}\int_{-\infty}^\infty \frac{d\omega}{\omega}
\int_{-\infty}^\infty
\frac{dk_y}{2\pi} g_{xx} (a,a;k_x=\omega/v,k_y,\omega),\label{friction1}
\ee
because the integration over $x'$ provides a $\delta$ function in 
$k_x-\omega/v$.  It will be noted that the only contributing
$k_x$ modes are those that keep pace with the particle, much like a surfer.
The Green's function appearing here can be written in terms of TE and TM parts,
indicated by $E$ and $H$ superscripts, respectively, in an arbitrary
background dielectric medium described by $\varepsilon(z;\omega)$, as
follows (for
details, see Refs.~\cite{Schwinger:1977pa,ce} and Appendix \ref{appa}):
\be
g_{xx}(z,z';k_x,k_y,\omega)=\frac{k_y^2}{k^2}\omega^2g^E(z,z';\kappa,\omega)
+\frac{k_x^2}{k^2}
\frac1{\varepsilon(z;\omega)}\frac1{\varepsilon(z';\omega)}
\partial_z\partial_{z'}g^H(z,z';\kappa,\omega).\label{teandtm}
\ee
Here $\kappa=\sqrt{k^2-\omega^2}$, and in the 
vacuum region $z>0$ above the conductor
\be
g^{E,H}(z,z';\kappa,\omega)=\frac1{2\kappa}\left(e^{-\kappa|z-z'|}
+r^{E,H}e^{-\kappa(z+z')}\right),\label{teandtmgf}
\ee
where the reflection coefficients at the interface of the uniform conductor 
with the vacuum are
\be
r^E=\frac{\kappa-\kappa'}{\kappa+\kappa'},\quad r^H=\frac{\kappa
-\kappa'/\varepsilon}{\kappa+\kappa'/\varepsilon},
\ee
in terms of $\kappa'=\sqrt{\kappa^2-\omega^2(\varepsilon(\omega)-1)}$.

The $1/i$ appearing in the frictional force (\ref{friction1}) is an instruction
 to take the
imaginary part. (Actually, the real part integrates to zero.)
 One might suppose that the propagation constant $\kappa$ would be complex, but
 due to the fact that $k_x=\omega/v$, that is entirely positive:
\be
\kappa^2=\frac{\omega^2}{\gamma^2-1}+k_y^2,
\ee
where $\gamma=(1-v^2)^{-1/2}$ is the usual relativistic dilation factor. 
Hence, only the parts of the Green's functions that are proportional to 
the reflection coefficients can contribute.

We find it convenient to introduce polar coordinates by defining the
 two-dimensional vector
\begin{subequations}
\be
\bm{\kappa}=\left(\frac\omega{\sqrt{\gamma^2-1}},k_y\right),\quad \bm{\kappa}^2
=\kappa^2,
\ee
so
\be
\omega=\sqrt{\gamma^2-1}\,\kappa\cos\theta,\quad k_y=\kappa\sin\theta.
\ee
\end{subequations}
 Then the frictional force can be written in the following general form
\be
F=-\frac{e^2}{8\pi^2}(\gamma^2-1)\int_0^\infty d\kappa\,\kappa\, 
e^{-2\kappa a}\int_0^{2\pi}
d\theta\frac{\cos\theta}{(\gamma^2-1)\cos^2\theta+1}[f^E(\kappa,
\theta;\gamma)+f^H(\kappa,\theta;\gamma)],\label{frictiongen}
\ee
where
\begin{subequations}
\be
f^E(\kappa,\theta;\gamma)=
2\sin^2\theta\,\Im\left[1+\sqrt{1-(\gamma^2-1)(\varepsilon-1)\cos^2\theta}
\right]^{-1},
\ee
and
\be
f^H(\kappa,\theta;\gamma)=
2\frac{\gamma^2}{\gamma^2-1}\Im\left[1+\frac1\varepsilon\sqrt{1
-(\gamma^2-1)(\varepsilon-1)\cos^2\theta}\right]^{-1}.\label{fhgen}
\ee
\end{subequations}

In the following, to be specific, we use the Drude model for the permittivity:
\be
\varepsilon(\omega)=1-\frac{\omega_p^2}{\omega^2+i\nu\omega},
\ee
where $\omega_p$ is the plasma frequency and $\nu$ is the damping parameter, 
assumed constant.  In terms of our polar 
variables, this translates
to
\be
\varepsilon-1=-\frac{\omega_p^2}{(\gamma^2-1)\kappa^2\cos^2\theta
+i\nu\kappa\sqrt{\gamma^2-1}\cos\theta}.
\ee
When we make specific numerical calculations, we can use approximate values for
 gold\footnote{More recent measurements by Olmon et al.~\cite{olmon}
give roughly consistent values: $\hbar\omega_p=8.5\pm0.5$ eV and
$\hbar\nu=0.050\pm0.011$ eV. We continue to use our nominal values
for illustrative purposes.} \cite{ba}:
\be
\hbar\omega_p=9.0 \,\mbox{eV},\quad\hbar \nu=0.035 \, \mbox{eV}.
\ee
(Again, for comparison with the quantum case, it is convenient to use the 
quantum-mechanical
energy conversion.  The conversion factor $\hbar c=2\times 10^{-5}
\mbox{eV\,cm}$ is useful.)

Let us adopt  dimensionless variables
\be
u=2\kappa a,\quad \alpha=2\omega_p a,\quad \beta=2\nu a,
\ee
to write the force as
\be
F=-\frac{e^2}{32\pi^2 a^2}\mathcal{F},
\ee
where
\be
\mathcal{F}=\mathcal{F}^E+\mathcal{F}^H=
(\gamma^2-1)\int_0^\infty du\,u \,e^{-u}\int_0^{2\pi}
\frac{ d\theta\cos\theta}
{(\gamma^2-1)\cos^2\theta+1}[f^E(u,\theta;\gamma;\alpha,\beta)
+f^H(u,\theta;\gamma;\alpha,\beta)].\label{friction2}
\ee

\section{TE contribution}
\label{sec:TE}
\begin{figure}
\includegraphics{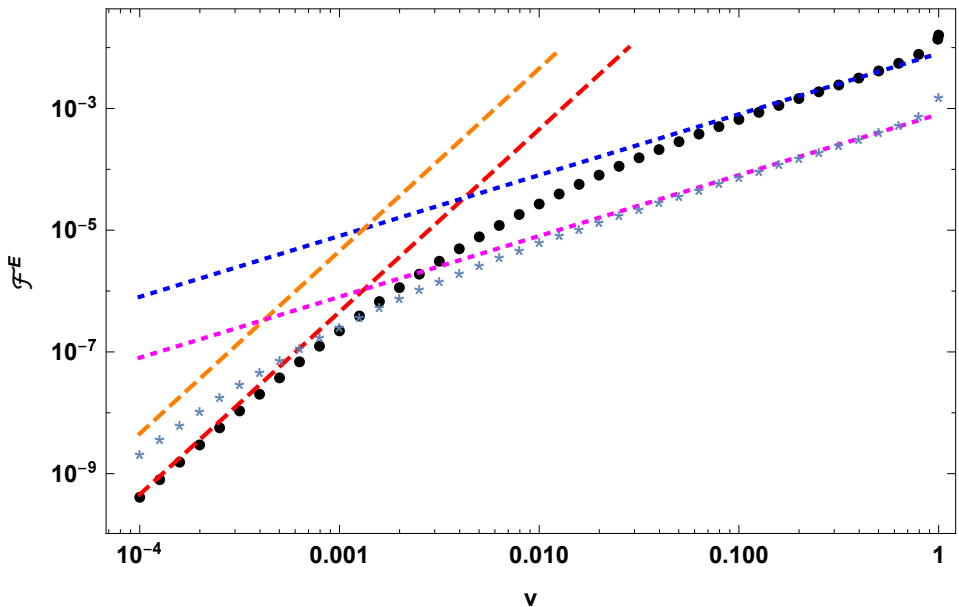}
\caption{\label{fig:allte} 
The numerically evaluated TE  frictional force 
(\ref{friction2}) [dots]
compared with the low velocity approximation (\ref{lowvfte}) [dashed red
line].  For larger velocities, not too close to the speed of light, the force
is well approximated by (\ref{modvte}) [short-dashed blue line].
In the ultrarelativistic limit, the TE friction approaches the value 
(\ref{ftelargeg}).  Here
we have chosen the separation distance of the particle from the plate to be 
$a=100\, \mbox{nm}$, where the Drude model should be approximately valid,
 so for gold, $\alpha=9.0$, $\beta=0.035$.
 Also shown, by stars, is the friction for one-tenth
the value of the dissipative parameter, $\beta=0.0035$, 
but with $\alpha$ unchanged, 
compared with the low-velocity [dashed orange] and intermediate velocity
[short-dashed magenta] approximations, which exhibits the enhancement at
low velocity and supression at high velocity caused by smaller dissipation.}
\end{figure}

Although it will turn out that the TE contribution is negligible compared to 
the TM part, it
is easier to analyze, so we start with that.  In the Drude model, the function 
$f^E$ is
\be
f^E(u,\theta;\gamma;\alpha,\beta)= 
2\sin^2\theta\,\Im\left[1+\sqrt{1+\frac{\alpha^2}{u^2}\left(1+i\frac{\beta}
{u\sqrt{\gamma^2-1}\cos\theta}\right)^{-1}}\right]^{-1}.\label{fte}
\ee
The exact numerical integration is a bit subtle and unstable; therefore,
we consider more tractable limits.
The nonrelativistic limit, $\gamma\to 1$, is straightforward:
\be
f^E(u,\theta;\gamma;\alpha,\beta)\to 
\frac{\alpha^2}{4u\beta}(\gamma^2-1)^{1/2}\cos\theta\sin^2
\theta,\quad v\ll\beta\ll 1,
\ee
which, when inserted into Eq.~(\ref{friction2}), yields
\be
\mathcal{F}^E=(\gamma^2-1)^{3/2}\frac\pi{16}\frac{\alpha^2}\beta=\frac\pi{16}
\frac{v^3\alpha^2}\beta,\quad v\ll \beta\ll 1.\label{lowvfte}
\ee
This agrees closely with the result of the direct numerical integration of the 
force for small
velocity of the charged particle, as seen in Fig.~\ref{fig:allte}.

For moderate velocities, the small $\beta$ expansion 
\be
f^E(u,\theta;\gamma;\alpha,\beta)\approx 
\frac\beta{\sqrt{\gamma^2-1}\cos\theta}\frac{\alpha^2
\sin^2\theta}{[u+\sqrt{u^2+\alpha^2}]^2} \frac1{\sqrt{u^2+\alpha^2}},
\quad \beta\ll v,
\label{smallzte}
\ee
reproduces the approximately  linear region  in Fig.~\ref{fig:allte}
 for intermediate velocities.   To see this, let
 $\gamma$ approach 1. The $\theta$ integral is just
$\int_0^{2\pi}d\theta\sin^2\theta=\pi$
and the remaining $u$ integral is
\be
I_E(\alpha)=\frac1{\alpha^2}\int_0^\infty du\,u\, e^{-u}
\frac{(\sqrt{u^2+\alpha^2}-u)^2}{\sqrt{u^2+\alpha^2}},
\ee
which can be written in terms of Struve and Bessel  functions as shown
in Appendix \ref{appc}. In this way we obtain
\be
\mathcal{F}^E\approx \pi \beta v  I_E(\alpha),\quad\beta\ll v\ll 1.
\label{modvte}
\ee
The agreement with numerical integration is good, as also shown in
Fig.~\ref{fig:allte}.

\begin{figure}
\includegraphics{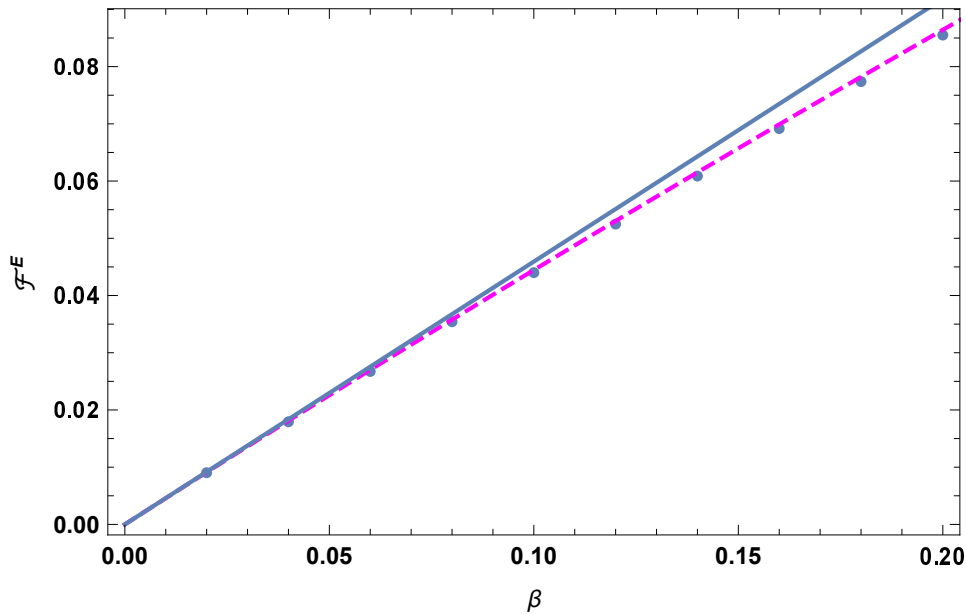}
\caption{\label{linte} 
The linear behavior (\ref{ftelargeg})
in the damping parameter $\beta$ of the 
ultrarelativistic TE friction [upper blue line], 
compared with exact data [dots] for $\alpha=9.0$, $\gamma=100$. 
For larger values of $\beta$ 
the data falls below the linear curve.  The curve that matches the
data well [magenta, dashed] 
is based on the more exact treatment (\ref{teexact}).
}
\end{figure}

Extracting the high-velocity limit is rather more subtle.  If we continue
to use the small $\beta$ expansion (\ref{smallzte}), 
we encounter  the $\theta$ integral 
\be
\int_0^{2\pi} d\theta\frac{\sin^2\theta}{\cos^2\theta+\frac1{\gamma^2-1}}
\approx2\pi\gamma, \quad \gamma\gg1.
\ee
  In terms of the function $I_E(\alpha)$, the TE friction in the 
$\gamma\to\infty$ limit approaches
\be
\mathcal{F}^E\sim 2\pi \beta  I_E(\alpha),
\label{ftelargeg}  
\ee 
twice the limit as $v\to 1$ of Eq.~(\ref{modvte}).
 In Fig.~\ref{linte} we show that this linear behavior in $\beta$ 
matches the exact integration quite well for low $\beta$.
Note, as further shown in Appendix \ref{appc}, Eq.~(\ref{Isabig}), that
the force tends to zero as $\omega_p\to\infty$, as we might expect for
a perfect conductor.

The astute reader might question the validity of the expansion 
(\ref{smallzte}) in powers
of the damping parameter $\beta$, since $\cos\theta=0$ is included in the
region of integration.  We can test this procedure in the
ultrarelativistic limit by breaking up
the $\theta$ integration into two intervals, $0<\theta<\theta_0$, 
$\theta_0<\theta
<\pi/2$, where $\pi/2-\theta_0\ll1$, but $\gamma\cos\theta_0\gg1$.  Then
the former interval is seen to give a contribution to the friction
which goes like $1/\gamma$ as $\gamma\to\infty$, while the latter can be
approximately written as
\be
\mathcal{F}^E\approx8\int_0^\infty du\,u \,e^{-u}\int_0^\infty
d\phi\frac\phi{\phi^2+1}\Im\left(1+\sqrt{
1+\frac{\alpha^2/u^2}{1+i\beta/(u\phi)}}\right)^{-1}.\label{teexact} 
\ee
Here $\phi=\gamma(\pi/2-\theta)$.
 Numerical integration of this is more 
stable than that of the original expression. Figure \ref{linte} displays the
result,
which matches the linear behavior for small $\beta$, and the exact data
for larger values of the damping.  
 Expanding this to first order in $\beta$, of course, yields
Eq.~(\ref{ftelargeg}).


\section{TM contribution}
\label{sec:TM}
\begin{figure}
\includegraphics{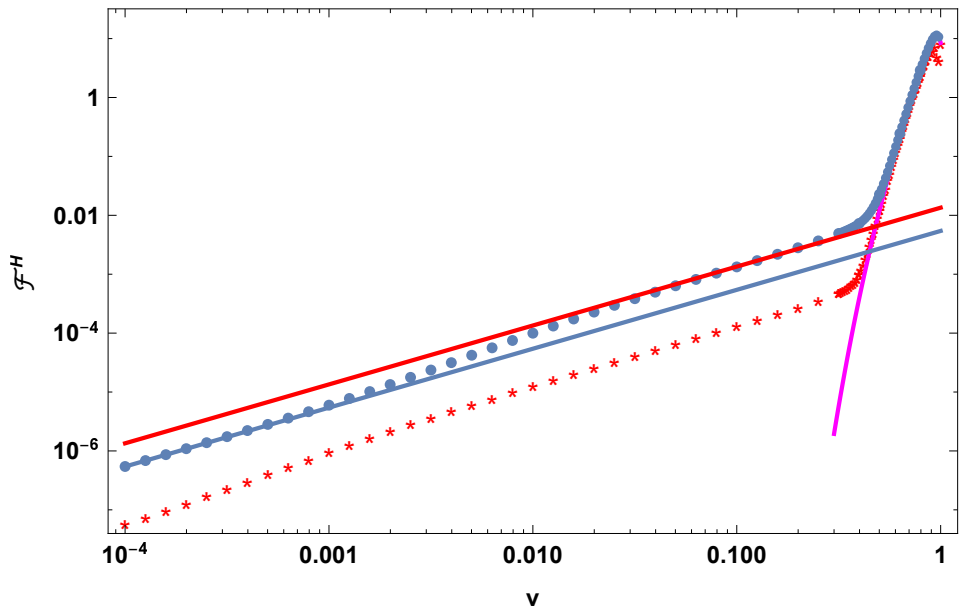}
\caption{\label{fig:alltm}  
The dependence of the TM frictional force on velocity [dots] compared with the 
low velocity approximation (\ref{lowvtm}) [lower, blue, straight line].  
Again the parameters are $\alpha=9.0$ and $\beta=0.035$.
 Agreement is very good for small velocities, $v\ll\beta$.  
For larger velocities,
the friction, computed numerically from Eq.~(\ref{friction2}), agrees well
with Eq.~(\ref{inttm}) [upper, red, straight line]  
for intermediate velocities, $\beta\ll v\ll 1$.
The high-velocity peak, for $v\sim 1$, is well reproduced by Eq.(\ref{beta0}) 
[magenta curve], which
 approaches the asymptotic value (\ref{ftmlim}), nonmonotonically,
with a maximum for $v<1$.  To demonstrate the effect of $\beta$, we
also plot the frictional force [red stars] 
for $\beta=0.0035$, one-tenth the value
above, but with the same plasma frequency parameter $\alpha$.  For low
and intermediate velocities, the friction is also reduced by a factor of 10, as
expected, but the high-velocity peak is unchanged.  (Instability of
the numerical integration for small $\beta$ is seen at velocities near
the speed of light.)}
\end{figure}

We turn now to the dominant TM contribution, which is, in general, rather more 
subtle.
The $v\to0$ limit is easy, since the leading contribution in the low $v$ limit 
is
\be
f^H(v,\omega)\to\frac{4\omega a}{v^2}\frac{\beta}{\alpha^2}.
\ee
Inserting this into the formula for the force (\ref{friction2}) we find  the 
low-velocity limit as given by Ref.~\cite{boyer1974}:
\be
\mathcal{F}^H=4\pi \frac{\beta}{\alpha^2}v,\quad\mbox{or}\quad
F^H\sim -\frac{e^2}{8\pi a^2}\frac{\beta v}{\alpha^2}=
-\frac{e^2}{16\pi}\frac{\nu v}{\omega_p^2}\frac1{a^3}, \quad v\ll \beta,
\label{lowvtm}
\ee
noting that the connection between the Drude-model parameters and the
Ohmic conductivity at zero frequency is $\sigma(0)=\omega_p^2/\nu$.
This is much larger than the $F^E$ contribution given in Eq.~(\ref{lowvfte}).  
We demonstrate
that this agrees with the exact numerical integration of the TM force in 
Fig.~\ref{fig:alltm}.

For somewhat higher velocities, $\beta\ll v\ll 1$, we can expand first in
$\beta/v$, and then in $v$, 
 and then writing $f^H=2\frac{\gamma^2}{\gamma^2-1}\Im\chi$, we
find
\be
\Im\chi=\frac{vu\beta\cos\theta}{\alpha^2}\sqrt{1+\frac{\alpha^2}{u^2}}
\left(1-\frac12\frac{\alpha^2/u^2}{1+\alpha^2/u^2}\right),
\ee
so when this is inserted into the formula (\ref{friction2})
 for $\mathcal{F}^H$ we obtain  (see Appendix \ref{appc})
\be
\mathcal{F}^H\approx 2\pi v\beta J_H(\alpha),\quad
J_H(\alpha)=\frac1{\alpha^2}\int_0^\infty du\,e^{-u}u
\sqrt{u^2+\alpha^2}\left(1-\frac12\frac{\alpha^2}{u^2+\alpha^2}\right),
\quad \beta\ll v\ll 1.
\label{inttm}
\ee
This agrees closely with the linear intermediate velocity region seen in 
Fig.~\ref{fig:alltm}.

\begin{figure}
\includegraphics{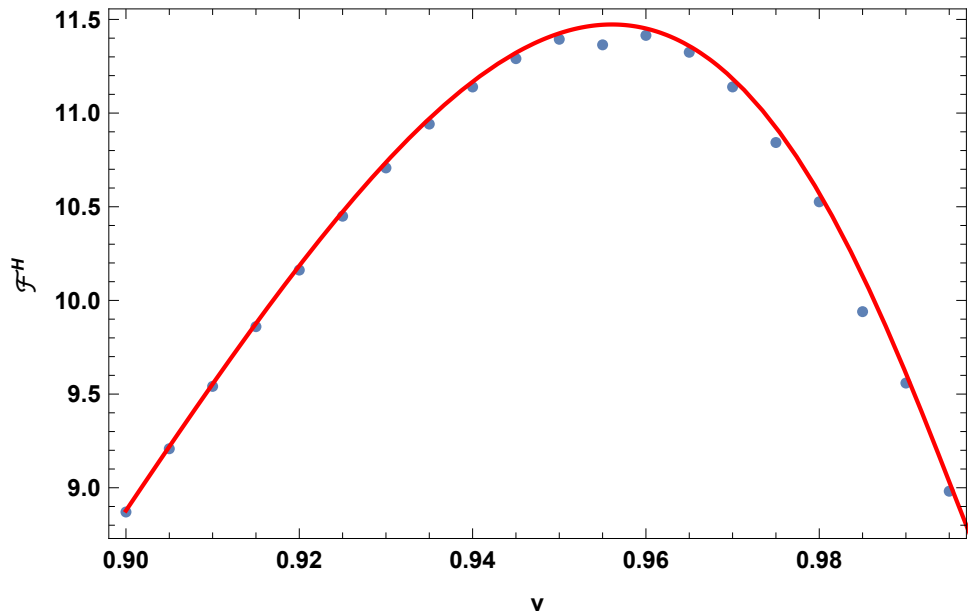}
\caption{\label{tmmax} The TM frictional force, obtained by numerical
integration, for values of $v$ for which $\mathcal{F}^H$ is maximum.  Again
we use nominal values of the plasma frequency and the damping parameter for 
gold, at a 100 nm separation, $\alpha=9.0$, $\beta=0.035$. The numerical
data [dots], which has some instability, is fit well by 
Eq.~(\ref{beta0}) [continuous curve]. The maximum value
is some 35\% larger than the limiting value given by Eq.~(\ref{ftmlim}).}
\end{figure}

\begin{figure}
\includegraphics{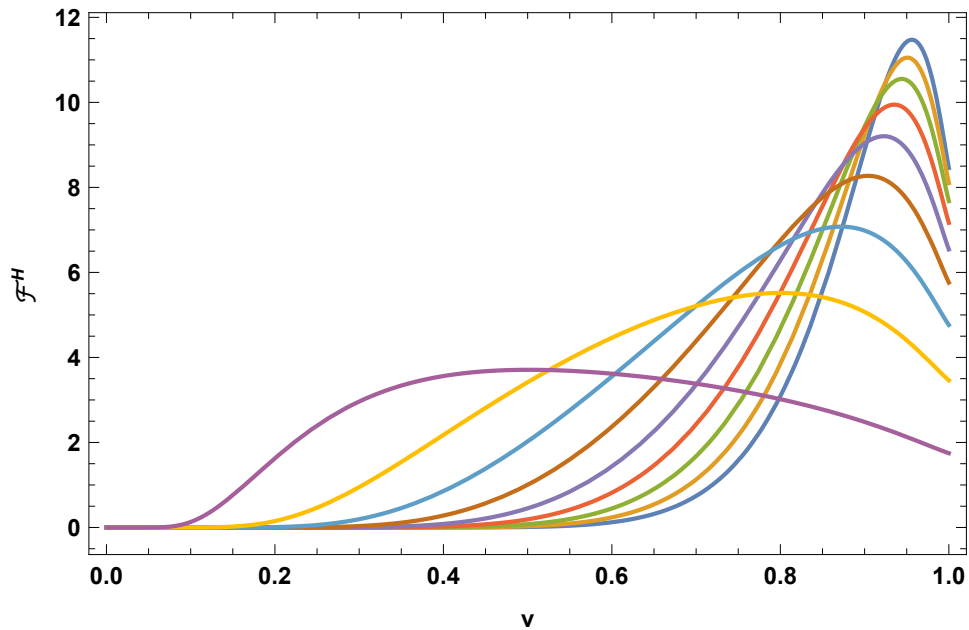}
\caption{\label{maxalpha} The function (\ref{beta0}) for various values
of $\alpha$, from $\alpha=1$  to $\alpha=9$  by
unit steps, which in all cases agrees with exact numerical data for the
TM (or total) frictional force.  The peak shifts to lower velocities as
$\alpha$ decreases, and the magnitude of the peak also decreases.}
\end{figure}

\begin{figure}
\includegraphics{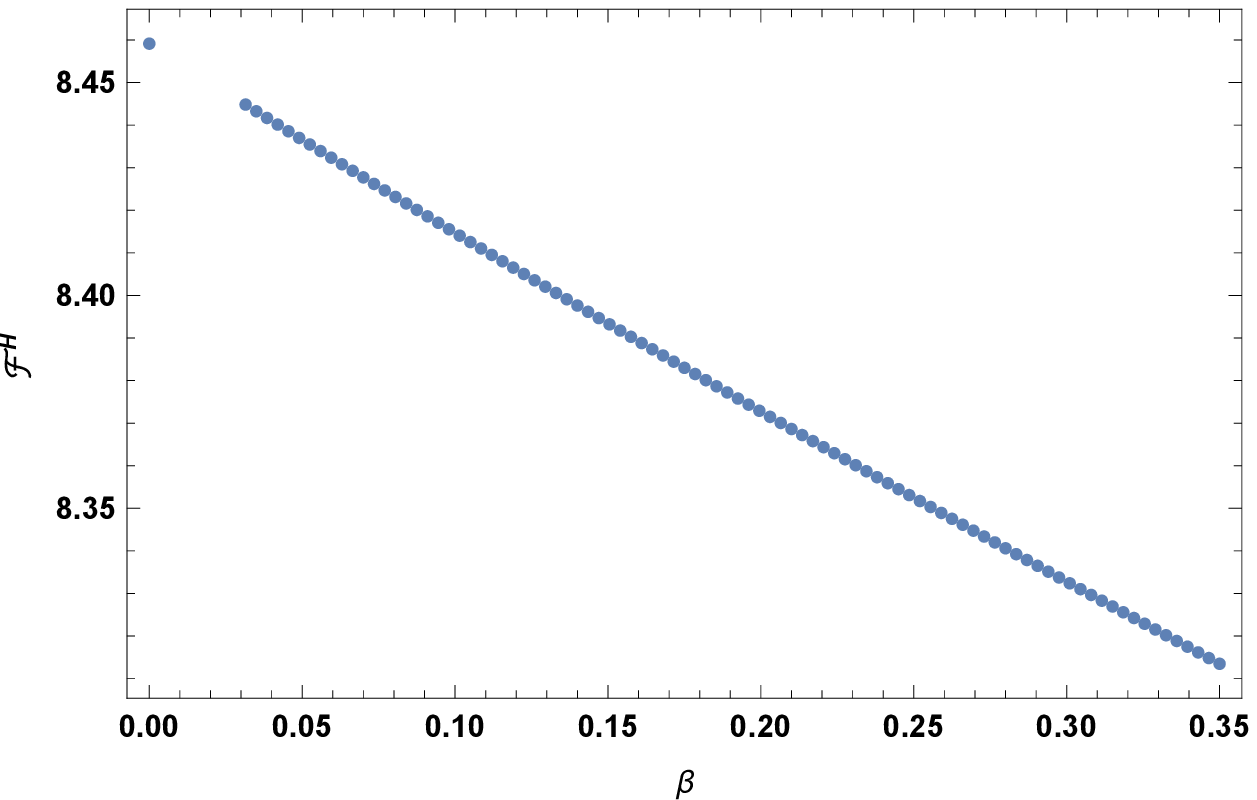}
\caption{\label{hitm} The TM frictional force in the $\gamma\to\infty$ limit
for $\alpha=9$ as a function of $\beta$.  Note there is a mild linear 
dependence on the damping parameter $\beta$, and that the force tends to a 
nonzero value for $\beta\to0+$. (The numerical integral becomes unstable 
for small values of $\beta$, but the limiting value (\ref{ftmlim}) at $\beta=0$
is shown.)}
\end{figure}

Turning to  higher velocities, we note that 
the expansion method in $\beta$
that worked well in the TE case fails.  This is because the force in this
case is no longer analytic in $\beta$; the integrand in the friction 
develops a singularity at $\beta=0$, for sufficiently high velocities.
We write Eq.~(\ref{fhgen}) exactly  as
\be
f^H=\frac{2\gamma^2}{\gamma^2-1}\Im \chi,\quad \mbox{where}\quad
\chi=\left\{1+\left[1-\frac{\alpha^2}{u^2\phi^2}
\frac1{1+i\beta/(u\phi)}\right]^{-1}
\sqrt{1+\frac{\alpha^2}{u^2}\frac1{1+i\beta/(u\phi)}}
\right\}^{-1},
\ee
with $\phi=\sqrt{\gamma^2-1}\cos\theta$.
The TM frictional force is then
\be
\mathcal{F}^H=
\frac{8\gamma^2}{\sqrt{\gamma^2-1}}\int_0^\infty du\,u\,e^{-u}\int_0^{\sqrt{
\gamma^2-1}} \frac{d\phi\,\phi}{\sqrt{\gamma^2-1-\phi^2}}\frac1{\phi^2+1}\Im
\chi.\label{tmex}
\ee
To get the relativistic behavior, as noted above, 
when $\beta=0$ the denominator in $f^H$ develops a pole at  $\phi=\phi_0$,
where
\be
\phi^2_0=\sqrt{1+\frac{\alpha^2}{u^2}}-1.
\ee
So as $\beta\to0$, we approximate $f^H$ by
\be
f^H\sim\frac{2\gamma^2}{\gamma^2-1}\Im\left[\frac{u\left(\phi^2-\frac{\alpha^2}
{u^2}\right)}{\lambda(\phi^2-\phi_0^2)+i\epsilon}\right],
\quad \lambda=u+\sqrt{u^2+\alpha^2}.\label{fheps}
\ee
Here $\epsilon$ is proportional to $\beta$, is always positive, 
 and for  $\alpha\gg u$
approaches  $\epsilon=\beta\sqrt{\alpha}/u^{3/2}$.
  Thus, for very small $\epsilon$,  
the imaginary part yields a $\delta$ function in $\phi$, which
lies in the region of the $\phi$ integration only if $\gamma^2>
\sqrt{1+\frac{\alpha^2}{u^2}}$. Thus, 
 we find that Eq.~(\ref{fheps}) implies in the limit $\beta\to0$ 
\be
\mathcal{F}^H\approx\frac{4\pi\gamma}{\sqrt{\gamma^2-1}}\frac1{\alpha^2}
\int_{\alpha/\sqrt{\gamma^4-1}}^\infty 
du\,u\,e^{-u}\frac{(u-\sqrt{u^2+\alpha^2})^2}
{\sqrt{1-\frac1{\gamma^2}\sqrt{1+\frac{\alpha^2}{u^2}}}}.
\label{beta0}
\ee
This agrees well with the exact $\mathcal{F}^H$ for high velocities, 
 is more stable numerically, and is
shown in Fig.~\ref{fig:alltm}.  The peak seen there is shown in more
detail in Fig.~\ref{tmmax}.
The dependence of this peak on $\alpha$ is shown in Fig.~\ref{maxalpha}.
From Eq.~(\ref{beta0}) we  obtain the limiting value for
$\gamma\to\infty$, $\beta\to0$:
\be
\mathcal{F}^H\to  4\pi  I_H(\alpha)=
\frac{4\pi}{\alpha^2}\int_0^\infty du\, u\, e^{-u}\left(
u-\sqrt{u^2+\alpha^2}\right)^2,\label{ftmlim}
\ee
 which remarkably is not zero.
See Appendix \ref{appc} for an explicit form for this integral.  It is plotted
in Fig.~\ref{Is}.
For small $\beta$ the ultrarelativistic limit exhibits a weak dependence
on the value of $\beta$, as shown in Fig.~\ref{hitm}.  This is
computed by taking the $\gamma\to\infty$ limit of Eq.~(\ref{tmex}),
and noting that only values of $\phi\alt \gamma$ are relevant:
\be
\mathcal{F}^H\sim 8\int_0^\infty du\,u\, e^{-u} \int_0^\infty 
\frac{d\phi \,\phi}{\phi^2+1}\Im \chi.
\ee
As shown in Appendix \ref{appc}, $I_H(\alpha)\to 1$ as 
$\alpha\to \infty$.
The difference between the dependencies of the
frictional force on the plasma frequency shown in Fig.~\ref{Is} is striking.  
This is correlated with the completely different dependence of the frictional
force on the dissipation parameter $\beta$.  Indeed, in the high-velocity,
large-plasma-frequency limit, 
\be
\frac{\mathcal{F}^E}{\mathcal{F}^H}\to \frac\beta{2\alpha},\quad \gamma\to
\infty,\, \alpha\to\infty.
\ee 

\section{Conclusions}
\label{sec:concl}
In this paper we have reconsidered classical friction between a charged
particle and an imperfectly conducting plate. We describe the latter
by the Drude model.  Only the nonrelativistic
regime had been considered previously, to our knowledge.  We examine both the
TE mode, which is quite negligible in practice, and the TM mode.  The low
velocity limit 
is very straightforward to analyze, but the limit of high velocities
(ultrarelativistic) is considerably more subtle.  We obtain results for all
velocities by a combination of analytic and numerical techniques.

The difference between the TE force seen in Fig.~\ref{fig:allte}
and the TM force seen in Fig.~\ref{fig:alltm}, is remarkable.
Not only is the value of the TE force typically orders of magnitude smaller,
but the TM force is nonmonotonic in the velocity.  It may seem surprising that
the maximum of the frictional force occurs for an intermediate value of
the velocity, as shown in more detail in Fig.~\ref{tmmax}, but this is due
to the appearance of a pole in the integrand for small damping.

How big are these effects, and might they be experimentally measurable?
We compare the largest value of the TM friction, $\mathcal{F}^H\approx 11.4$,
with our parameters,
from Fig.~\ref{fig:alltm}, with the force on a static 
charged particle  next to a conducting
plate, $F_c=-e^2/(16\pi a^2)$.  For our nominal values $\alpha=9$, 
$\beta=0.035$,
the ratio is maximum at about 0.96 times the speed of light:
\be
\frac{F}{F_c}\le 1.81.
\ee
This  should be readily observable.  This ratio drops to 
about 1.33 for an ultrarelativistic charged particle.
 
\acknowledgments{We thank our collaborators, Prachi Parashar, Steve Fulling,
 Hannah Day, Aaron Swanson, and Dylan DelCol 
for many helpful comments. We thank an anonymous referee for
extremely insightful comments.  This work was supported in part
by a grant from the US National Science Foundation, grant number 1707511.}

\appendix
\section{Electromagnetic Green's Function}
\label{appa}
Maxwell's equations in a medium characterized by position- and frequency-%
dependent permittivity $\varepsilon$ and permeability $\mu$ yield the
wave equation for the electric field
\be
\bm{\nabla}\times\frac1\mu\bm{\nabla}\times \mathbf{E}-\omega^2
\varepsilon\mathbf{E}=i\omega \mathbf{j},
\ee
where $\mathbf{E}=\mathbf{E}(\mathbf{r};\omega)$,
$\mathbf{j}=\mathbf{j}(\mathbf{r};\omega)$.
The electromagnetic Green's dyadic satisfies a similar equation,
\be
\left[\frac1{\omega^2}\bm{\nabla}\times\frac1\mu\bm{\nabla}\times-\varepsilon
\right]\bm{\Gamma}(\mathbf{r,r'};\omega)=\bm{1}\delta(\mathbf{r-r'}).
\ee
From this Eq.~(\ref{egf}) immediately follows.

For the planar geometry we are considering for the dielectric slab, the Green's
dyadic possesses translational invariance in the plane of the slab, the $x$-$y$
plane, so we have the Fourier representation (\ref{tft}),
where, in a coordinate system in which $\mathbf{k_\perp}$ lies in the $x$ 
direction, $\mathbf{g}$ breaks up into block-diagonal form:
\be
\mathbf{g}(z,z';\mathbf{k_\perp},\omega)=\left(\begin{array}{ccc}
\frac1\varepsilon \partial_z\frac1{\varepsilon'}\partial_{z'}g^H
-\frac1\varepsilon\delta(z-z')&0&\frac{ik}{\varepsilon\varepsilon'}\partial_z
g^H\\
0&\omega^2 g^E&0\\
-\frac{ik}{\varepsilon\varepsilon'}\partial_{z'}g^H&0&\frac{k^2}{\varepsilon
\varepsilon'}g^H-\frac1\varepsilon\delta(z-z')
\end{array}\right).\label{gee}
\ee
Here $\varepsilon=\varepsilon(z)$, $\varepsilon'=\varepsilon(z')$, 
$k=|\mathbf{k}_\perp|$, and
$g^E(z,z')$, $g^H(z,z')$ are the transverse electric and transverse magnetic
Green's functions, which satisfy [in a general medium with both permittivity
$\varepsilon=\varepsilon(z,\omega)$ and permeability $\mu=\mu(z,\omega)$]
\begin{subequations}
\bea
\left(-\frac{\partial}{\partial z}\frac1\mu\frac\partial{\partial z}+\frac{k^2}
\mu-\omega^2\varepsilon\right)g^E(z,z')&=&\delta(z-z'),\\
\left(-\frac{\partial}{\partial z}\frac1\varepsilon\frac\partial{\partial z}
+\frac{k^2}\varepsilon-\omega^2\mu\right)g^H(z,z')&=&\delta(z-z').
\eea
\end{subequations}
For the case of a homogeneous dielectric slab extending over the half-space
$z<0$, the solution of these equations for $z>0$ is given in terms of 
reflection coefficients by Eq.~(\ref{teandtmgf}), as shown in textbooks, 
for example Ref.~\cite{ce}.  The $\delta$-function terms in Eq.~(\ref{gee})
are to be omitted, as ``contact terms,'' because we always take the
coincident point {\it limit}.

For the application here, we have to remove the restriction that 
$\mathbf{k_\perp}$ lie along the $x$ axis, which we do by the orthogonal
transformation
\be
\mathbf{\tilde g}=\mathbf{O}\mathbf{g}\mathbf{O}^T, \quad \mathbf{O}=\left(
\begin{array}{ccc}
\frac{k_x}k&-\frac{k_y}k&0\\
\frac{k_y}k&\frac{k_x}k&0\\
0&0&1\end{array}\right).
\ee
Equation (\ref{teandtm}) now follows.

\section{Evaluation of Integrals}
\label{appc}

It is straightforward to show that the integrals occurring in the 
 ultrarelativistic limit ($v\to 1$) for $\mathcal{F}^E$ 
\begin{subequations}
\begin{equation}
I_E(\alpha)=\alpha^2\int_0^{\infty}du\,e^{-u} \frac{u}{\sqrt{u^2+\alpha^2}
\left(u+\sqrt{u^2+\alpha^2}\right)^{\!2}}\end{equation}
and in the ultrarelativistic  limit for $\mathcal{F}^H$  
\begin{equation}
I_H(\alpha)=\alpha^2 \int_0^{\infty}du\,e^{-u} 
\frac{u}{\left(u+\sqrt{u^2+\alpha^2}\right)^{\!2}}
\end{equation}
\end{subequations}
may be expressed in terms of Struve and Bessel functions\footnote{
It is interesting to note that the general formulas given, for 
example in Ref.~\cite{tomassone1997} for the nonrelativistic case,
involve the same combination of Struve and Bessel functions; in
that case $\mathbf{H}_0(\xi)-Y_0(\xi)$, where $\xi=\pi\alpha^2/(\beta v)$.
However, beyond the leading low-velocity term (\ref{lowvtm}), 
the corrections they give are very small, and do not describe the
deviation from linearity that we see, for example, in Fig.~\ref{fig:alltm}.}
 by using \cite{gs}
\begin{equation}
\int_0^{\infty}du\,e^{-u}\left(u^2+\alpha^2\right)^{\nu-1}
=\frac{\sqrt{\pi}}{2}\,(2\alpha)^{\nu-\frac12}\,\Gamma(\nu)
\left[\mathbf{H}_{\nu-\frac12}(\alpha)-Y_{\nu-\frac12}(\alpha)\right].
\end{equation}
Thus,
\begin{subequations}
\begin{eqnarray}
I_E(\alpha)&=&\frac{1}{\alpha^2}\int_0^{\infty}du\,e^{-u}\,\frac{u
\left(u-\sqrt{u^2+\alpha^2}\right)^{\!2}}{\sqrt{u^2+\alpha^2}}\nn\\
&=&\frac{1}{\alpha^2}\int_0^{\infty}du\,e^{-u}\,\frac{\left(-\alpha^2 u+2u(u^2
+\alpha^2) -2u^2\sqrt{u^2+\alpha^2}\right)}{\sqrt{u^2+\alpha^2}}\nn\\
&=&\frac{1}{\alpha^2}\int_0^{\infty}du\,e^{-u}\left(-\alpha^2\,\frac{d}{du}
\sqrt{u^2+\alpha^2}+\frac23\,\frac{d}{du}\left(u^2+\alpha^2\right)^{\!\frac32}
-2u^2\right)\nn\\
&=&\frac{1}{\alpha^2} \left[\frac{\alpha^3}{3}+\int_0^{\infty}du\,e^{-u}
\left(-\alpha^2 \sqrt{u^2+\alpha^2}+\frac23 \left(u^2+\alpha^2\right)^{\!
\frac32}-2u^2\right)\right]\nn\\
&=&-\frac{4}{\alpha^2}+\frac{\alpha}{3}-\frac{\pi\alpha}{2}
\left[\mathbf{H}_1(\alpha)-Y_1(\alpha)\right]+\pi
\left[\mathbf{H}_2(\alpha)-Y_2(\alpha)\right].
\end{eqnarray}
Likewise,
\begin{eqnarray}
I_H(\alpha)&=&\frac{1}{\alpha^2}\int_0^{\infty}du\,e^{-u}\,u 
\left(u-\sqrt{u^2+\alpha^2}\right)^{\!2}\nn\\
&=&\frac{1}{\alpha^2}\int_0^{\infty}du\,e^{-u}\left(\alpha^2 u+2 u^3 +2\alpha^2
\sqrt{u^2+\alpha^2}-2 \left(u^2+\alpha^2\right)^{\!\frac32}\right)\nn\\
&=&\frac{12}{\alpha^2}+1+\pi\alpha\left[\mathbf{H}_1
(\alpha)-Y_1(\alpha)\right]-3\pi\left[\mathbf{H}_2(\alpha)
-Y_2(\alpha)\right].
\end{eqnarray}
\end{subequations}

\begin{figure}
\includegraphics{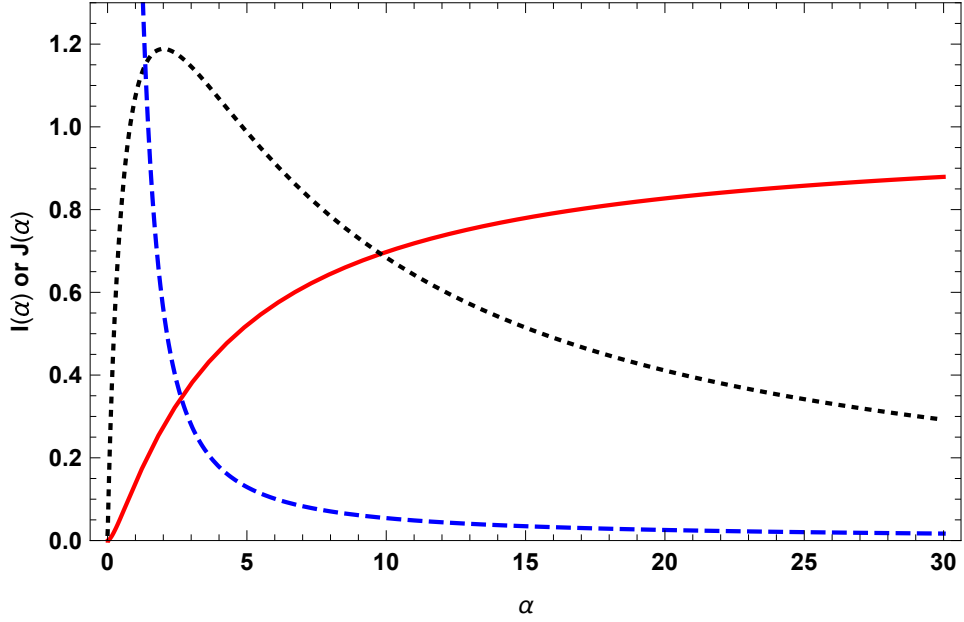}
\caption{\label{Is} Behavior of the integrals $ I_{H,E}$ as  functions
of the plasma frequency parameter $\alpha$.  The TE contribution 
[black, dotted] $I_E$ is multiplied by a factor of 10, so the
two functions may be shown on the same graph. The TM integral $I_H$
is shown by the solid red curve.  These functions describe
the high-velocity limit of the frictional force, according to 
Eqs.~(\ref{ftelargeg}) and (\ref{ftmlim}). 
Also plotted is the function
$J_H(\alpha)$ [dashed, blue, curve], 
which describes the intermediate-velocity dependence of the
TM frictional force, according to Eq.~(\ref{inttm}).}
\end{figure}

For small values of $\alpha$, standard expansions of $\mathbf{H}_n(\alpha)$ 
and $Y_n(\alpha)$ may be used to evaluate these functions:
\begin{subequations}
\bea
 I_E&\sim&\frac\alpha3-\frac{\alpha^2}{16}
\left(1-4\gamma_E-4\ln\frac\alpha2\right)-\frac1{5}\alpha^3+\dots,\\
I_H&\sim& -\frac{\alpha^2}{16}\left(1+4\gamma_E+4\ln\frac\alpha2\right)
+\frac4{15}\alpha^3+\dots,
\eea
\end{subequations}
for $\alpha\ll 1$,
in terms of Euler's constant, $\gamma_E=0.57721\dots$. For large values of 
$\alpha$, the following asymptotic expansion \cite{gs} may be employed:
\begin{equation}
\mathbf{H}_n(\alpha)-Y_n(\alpha)=\frac{1}{\pi}\sum_{m=0}^{p-1}\frac{\Gamma
\left(m+\frac12\right)}{\Gamma\left(n+\frac12-m\right)}\left(\frac{\alpha}{2}
\right)^{\!n-1-2m}+O\left(\alpha^{n-1-2p}\right).
\end{equation}
It follows that 
\be
 I_E\sim \frac1\alpha-\frac4{\alpha^2},\quad
I_H\sim 1-\frac4\alpha,
\quad \mbox{as} \quad \alpha\to\infty.\label{Isabig}
\ee
These functions are plotted in Fig.~\ref{Is}.

$I_E(\alpha)$ also describes the behavior of $\mathcal{F}^E$ for intermediate
velocities, according to Eq.~(\ref{modvte}), while the corresponding
intermediate-velocity behavior of $\mathcal{F}^H$ is given by Eq.~(\ref{inttm})
in terms of $J_H(\alpha)$, where
\bea
J_H(\alpha)&=&\frac1{\alpha^2}\int_0^\infty du\,e^{-u}u
\sqrt{u^2+\alpha^2}\left(1-\frac12\frac{\alpha^2}{u^2+\alpha^2}\right)\nn\\
&=&\frac{\alpha}6+\frac{\pi}2\left(\mathbf{H}_2(\alpha)-Y_2
(\alpha)\right)-\frac{\alpha\pi}4\left(\mathbf{H}_1(\alpha)-Y_1(\alpha)\right).
\eea
This is also plotted in Fig.~\ref{Is}.
The behaviors for large and small values of $\alpha$ are
\begin{subequations}
\bea
J_H(\alpha)&\sim& \frac2{\alpha^2}+\frac\alpha6+\frac{\alpha^2}{32}
\left(-1+4\gamma_E+4\ln\frac\alpha2\right),\quad \alpha\ll 1,\\
&\sim&\frac1{2\alpha}+\frac9{2\alpha^3},\quad \alpha\gg 1.
\eea
\end{subequations}

\section{Vavilov-\v{C}erenkov Radiation}
\label{sec:vc}

To illustrate the further utility of our Green's function approach, we
apply Eqs.~(\ref{friction1}) and (\ref{teandtm}) 
to the situation of a charged particle moving
through a homogeneous nondissipative
dielectric material faster than the speed of light in the medium,
$1/n=1/\sqrt{\varepsilon}$.  In this case we will disregard dissipation in the
material, setting $\nu=0$; the imaginary part comes 
from the region of frequencies
where $v>1/n(\omega)$.  The TE part of the drag on the particle is given
by
\be
F^E=-\frac{e^2}{2\pi}\int\frac{d\omega}\omega \int_{-\infty}^\infty \frac{dk_y}
{2\pi}\frac{\omega^2}{k_y^2+\omega^2/v^2}k_y^2\,\Im\frac1{2\kappa'},
\ee
since we now only have the bulk (first) term in Eq.~(\ref{teandtmgf}), 
except that the particle is in the medium, so $1/(2\kappa)\to 1/(2\kappa')$.
The branch line is chosen to run between the two branch points, 
where $k_y^2=n(\omega)^2\omega^2[1-1/(n(\omega)^2 v^2)]$,
on the real $k_y$ axis, 
The subtlety is the sign of the imaginary part.  This is resolved
by noting that  the retarded
Green's function must have singularities only in the lower-half $\omega$ plane,
which is consistent with the requirement that, in the case of infinitesimal
damping, $n(\omega)^2\omega^2$ has an imaginary part $\epsilon\sgn(\omega)$,
with $\epsilon\to0+$.  Therefore, the $k_y$ integration passes below the
branch line for $\omega>0$, and above for $\omega<0$.  In dimensionless form,
that integral then is
\be
\int_{-1}^{1} dx\frac{x^2}{x^2+a^2}\frac{\sgn{\omega}}{\sqrt{1-x^2}}=
\pi\sgn(\omega)\left(1-\frac{a}{\sqrt{1+a^2}}\right),
\ee
with $a=\left(vn(\omega)\sqrt{1-1/(vn(\omega))^2}\right)^{-1}$, so that
the above integral is simply $\sgn(\omega)\pi[1-1/(vn(\omega))]$.  
The resulting drag force due to \v{C}erenkov radiation is
\be
F^E=-\frac{e^2}{8\pi}\int d\omega\,|\omega|\left(1-\frac1{n(\omega)v}\right),
\ee
where the integral is over the region where $n(\omega)>1/v$.  

The TM contribution to the drag force is 
\be
F^H=\frac{e^2}{2\pi}\int\frac{d\omega}\omega \int_{-\infty}^\infty \frac{dk_y}
{2\pi}\frac{\omega^2/v^2}{k_y^2+\omega^2/v^2}\Im\left[\frac{\kappa^{\prime 2}}
{\varepsilon(\omega)^2 }\frac{\varepsilon(\omega)}{2\kappa'}\right],
\ee
because except in the
exponent, the TM Green's function is obtained from
that for TE by the replacement $\kappa'\to\kappa'/\varepsilon$.
After doing the $k_y$ integral as above, which now is
\be
-\sgn(\omega)\int_{-1}^1 dx\frac{\sqrt{1-x^2}}{x^2+a^2}=\pi\sgn(\omega)
\left(1-vn(\omega)\right),
\ee
we have
\be
F^H=-\frac{e^2}{8\pi}\int d\omega\,|\omega| \frac1{n(\omega)v}\left(1-\frac1
{n(\omega)v}\right).
\ee

Adding the two modes together,
\be
F=F^E+F^H=-\frac{e^2}{4\pi}\int d\omega\,\omega 
\left(1-\frac{1}{n(\omega)^2v^2}\right),
\ee
where now the integration is over
{\it  positive\/} frequencies for which the speed
of the particle  exceeds that of light in the medium, $1/n(\omega)$.
This formula exactly coincides with the energy loss rate found in Eq.~(36.19) 
of Ref.~\cite{ce} due to the energy radiated by Vavilov-\v{C}erenkov effect.
(Note, Gaussian units were used there, and $e_{\rm HL}^2=4\pi e_{\rm G}^2$.)


\end{document}